\documentclass[12pt,titlepage]{article}

\usepackage{setspace} 
\usepackage{amsmath}
\usepackage{graphicx}
\usepackage[round,numbers,sort&compress]{natbib} 

\newcommand{\bit}{\begin{itemize}}

\newcommand{\eit}{\end{itemize}}
\newcommand{\beq}{\begin{equation}}
\newcommand{\beqn}{\begin{eqnarray}}
\newcommand{\eeq}{\end{equation}}
\newcommand{\eeqn}{\end{eqnarray}}
\title{Robust entrainment of circadian oscillators requires specific phase
  response curves} 

\author{Benjamin~Pfeuty\\
        Laboratoire de Physique des Lasers, Atomes, et Mol\'ecules, CNRS,
        UMR8523 \\ 
	Interdisplinary Research Institute, CNRS, USR3078 \\
	Universit\'e Lille 1, F-59655 
	Villeneuve d'Ascq, France
	\and Quentin Thommen \\
	Laboratoire de Physique des Lasers, Atomes, et Mol\'ecules, CNRS,
        UMR8523 \\ 
	Interdisplinary Research Institute, CNRS, USR3078 \\
	Universit\'e Lille 1, F-59655 
	Villeneuve d'Ascq, France\\
	\and Marc Lefranc \\
	Laboratoire de Physique des Lasers, Atomes, et Mol\'ecules, CNRS,
        UMR8523 \\ 
	Interdisplinary Research Institute, CNRS, USR3078 \\
	Universit\'e Lille 1, F-59655 
	Villeneuve d'Ascq, France
}

\date{}

\begin{document}

\maketitle

\abstract{The circadian clocks keeping time of day in many living organisms
  rely on self-sustained biochemical oscillations which can be entrained 
  by external cues, such as light, to the $24$-hour cycle induced by
Earth rotation.
However, environmental cues are unreliable due to the variability of habitats,
weather conditions or cue-sensing mechanisms among individuals. 
A tempting hypothesis is that circadian
clocks have evolved so as to be robust to fluctuations in daylight or
other cues when entrained by the day/night cycle. 

To test this hypothesis, we analyze the synchronization behavior of weakly and
periodically forced oscillators in terms of their 
phase response curve (PRC), which measures phase changes induced by a
perturbation applied at different phases. 
We establish a general relationship between, on the one side, the robustness
of key entrainment properties such as stability and phase shift and, on
the other side, the shape of the PRC as characterized by a specific
curvature or the existence of a dead zone.
This result can be applied to computational models of circadian clocks 
where it accounts for the disparate robustness properties of various
forcing schemes.
Finally, the analysis of PRCs measured
experimentally in several organisms strongly suggests a case of convergent
evolution toward an optimal strategy for maintaining a clock that is accurate 
and robust to environmental fluctuations.}

\section*{INTRODUCTION}%

Circadian entrainment is the process by which a biological clock
with a period of approximately 24 hours is synchronized to environmental
cycles associated with Earth rotation. 
A stable and precise phase relationship between internal and external
times is vital for organisms that need to timely anticipate dawn or dusk in
order to coordinate their physiology to diurnal environmental changes.
For instance, a precise clock in phototrophic organisms like cyanobacteria or
plants has been shown to optimize cell growth and fitness
\cite{Woelfle04,Dodd05,Graf10}. 

However, circadian clock precision is challenged by many sources of intrinsic
and extrinsic variability, which could dramatically affect key properties of
endogenous biochemical oscillations. 
A growing interest has therefore been devoted to investigate 
how the period and amplitude of these oscillations are impacted
by genetic mutations \cite{Price98,Cheng01}, molecular noise
\cite{Barkai00,Gonze02,Mihal04,Wang08},  
or contextual variability \cite{Mihal04,Dibner09},
which led to advocate the existence of design principles ensuring robust
oscillatory behavior in circadian clocks \cite{Rand04,Stelling04,Wagner05}.

Nevertheless, a robust endogenous clock does not guarantee by itself a precise
phase relationship with the day-night cycle, since the 
environmental cues associated with the diurnal changes are also highly
fluctuating.
The daylight intensity and quality sensed by an organism depend on
various environmental factors such as meteorological conditions, shade
habitats or, for marine organisms, the distance to sea surface and water
turbidity \cite{Stramska92,Graham03}. 
In addition, variations of the behavior and light-sensing abilities of
individuals can also alter the light signal reaching their core molecular
clock. 
Thus, individuals of the same species living in different places or times
can experience a wide spectrum of light intensities over several orders of
magnitude, raising the question of whether and how specific robustness
strategies are implemented in their clock architectures to maintain a precise
synchronization despite unreliable environmental cues.
Although the importance of this problem was noted some time ago
\cite{Beersma99},  the robustness of circadian clocks to daylight fluctuations
and how this constraint shapes their molecular architecture have been little
studied until very recently \cite{Troein09,Thommen10}. 
These computational studies have revealed disparate robustness capabilities
depending on the clock model, which remains to be explained in a comprehensive
approach.

A natural theoretical framework to address this issue is the dynamical system
theory of synchronization \cite{Pikovsky00,Winfree01}.
In forced oscillatory systems, a phase-locked state specified by a stable
phase shift between the external and the internal phases closely depends on
forcing features such as amplitude, profile or period. 
Changing these forcing properties can alter the phase shift or 
 induce complex dynamical regimes (e.g., quasi-periodic, period-doubled,
 or chaotic),
 which are both undesirable for a functional circadian clock.
Thus, the question of whether a  circadian clock is robust to daylight
fluctuations can be generalized to the problem of how an oscillator
can maintain a stable synchronization and phase relationship with a forcing
cycle that exhibits significant variability.
Assuming a weak forcing allows one to drastically simplify
the theoretical analysis by utilizing
perturbation theory in the 
vicinity of a periodic orbit \cite{Winfree67,Kuramoto84,Kramer84}, 
in which the infinitesimal phase response
curve (IPRC) of a weakly coupled oscillator provides important
information on its synchronization. 

In this paper, we first lay out the theoretical approach within which
two
quantities measuring robustness of the entrained state with respect to
forcing fluctuations are defined.
The phase approximation in the weak forcing limit allows us to
identify the geometric properties of phase response curves that 
contribute to the robustness and precision of the clock phase in presence
of fluctuations in the forcing amplitude.   
The general  criteria obtained are shown to explain and predict the robustness
properties of biologically-based circadian models.
Finally, the analysis of PRCs that have been measured experimentally in several
organisms supports the idea that living organisms have evolved salient
strategies to maintain an accurate clock that is robust to daylight
fluctuations.

\section*{RESULTS}%

\subsubsection*{Circadian clock entrainment and phase approximation}

The entrainment of circadian clocks by cyclic environmental changes
is a paradigmatic example of an unidirectional synchronization
process. 
A biochemical implementation of this process in a living cell involves
a network 
of genes and proteins interacting with each other, whose
 temporal evolution is usually modeled 
by a set of ordinary differential equations:
\begin{equation} 
\frac{d\mathbf{X(t)}}{dt}=\mathbf{F}(\mathbf{X}(t),\mathbf{p_0}+\mathbf{dp}(t))    
\label{dyn}
\eeq
where the components of vector $\mathbf{X}$ are the
concentrations of the molecular actors interacting according to the
biochemical kinetics $\mathbf{F}$,
derived from the law of mass action.

Synchronization of the circadian clock to the diurnal cycle requires that some
parameters $p_i$ are sensitive to 
temporal changes of light induced by the cycle, so that they differ from their
value in the dark $(p_0)_i$ by
\begin{equation}        
dp_i(t)=\epsilon \, L(u)\,v_i\,(p_0)_i
\label{forcing}
\end{equation}
where the perceived light intensity is described by an amplitude
$\epsilon$ and an a normalized temporal profile $L$ which depends on
$u=t-t_d$, with $u\in[0,T]$, $t_d$ and $T$ corresponding 
to dawn time and the $24$-hour day length, respectively. 
The relative sensitivities of parameters to light are given by the
vector $\mathbf{v}$, 
whose is normalized ($||\mathbf{v}||=1$) to ensure that
$\epsilon$ is a relative modulation amplitude.  
In day/night entrainment conditions, the light sensed by the organism is
assumed to be restricted to daytime of duration $\tau_D$ ($L(u)=0$ for
$u\in[\tau_D,T]$).  
In this work, an important point is that $\epsilon$ and $L$ can differ between
individuals depending on environmental or physiological context. 

The existence of endogenous circadian oscillations requires that
Eqs. \ref{dyn} parameterized by $\mathbf{p}_0$ in the absence of light have an
asymptotically stable limit cycle solution $\mathbf{X}_{0}$ characterized by a
free-running period (FRP) $T_{0}$.
The light-dependent perturbation deviates the circadian oscillator from its
free limit cycle trajectory during daytime.
If the amplitude deviation is not too large, the stability of the limit cycle
ensures that the deviation decays in the absence of perturbation
(during night), with the only memory of the past perturbation being a
residual phase change.
Thus, the Poincar\'e map of the dynamical system described by Eqs. \ref{dyn}
can be approximated by the following unidimensional map \cite{Rand04}: 
\begin{equation}
\phi_{n+1} = G(\phi_n)= \phi_n - \gamma + V(\phi_n,\epsilon) 
\label{map}
\end{equation}
where $\phi_n \in [0,T]$ is the oscillator phase in units of circadian hours
(abbreviated as ch) \cite{Pittendrigh64} at which the light-dependent
perturbation is switched on during the  $n$-th day (i.e., the oscillator phase
at dawn).
$\gamma=(T_{0}-T)T/T_{0}$ is the phase difference associated with the
period mismatch between the forcing and endogenous periods. 
$V(\phi_n,\epsilon)$ is the phase change induced by
the perturbation given by Eq.~\ref{forcing} applied at oscillator phase
$\phi_n$, and is known 
as a phase response curve (PRC) in the litterature \cite{Winfree01}.  
The mapping $G$ has a stable fixed point if there exists a phase shift
$\phi^*$ that
satisfies:
\begin{eqnarray}
\left\{ \begin{array}{ll}
V(\phi^*,\epsilon)=\gamma\\
-2<\partial_{\phi} V(\phi^*,\epsilon)<0
\end{array} \right.
\label{stabcond}
\end{eqnarray}
This phase-locked state is usually termed $1:1$ synchronization (or
entrainment) 
state. The  stability coefficient $\partial_{\phi}
V(\phi^*,\epsilon)$ (also noted $V'(\phi^*)$) characterizes how fast
 fluctuations around $\phi^*$ decay.

Due to the transversal stability of the limit cycle and the neutral stability
of the phase, one can assume that small perturbations mainly induce phase
deviations, which comprehensively  describe the effect on oscillator dynamics. 
In the framework of the so-called phase approximation  \cite{Kuramoto84},
the PRC $V(\phi)$ can then be derived from an impulse
infinitesimal PRC (i-IPRC) $Z(u)$ according to a convolution integral, with
the important property that it scales linearly with light stimulus amplitude
\cite{Rand04}:
\begin{equation}
V(\phi,\epsilon) = \epsilon \int_0^{\tau_D}
Z(u+\phi)L(u) du = \epsilon \, W(\phi) 
\label{prop}
\end{equation}
Besides the classic i-IPRC $Z(\phi)$, Eq. \ref{prop} also
introduces the d-IPRC $W(\phi)$ that describes the phase response of  
the oscillator to day-like light perturbations of duration $\tau_D$, temporal
profile $L(u)$ and vanishingly small amplitude $\epsilon$.

\subsubsection*{Two distinct metrics to measure clock entrainment
  robustness} 

The daylight intensities sensed by a living organism depend on environmental
factors, habitat or cloud cover, which can vary in time and space.
They can also vary due to the temporal or inter-individual variability of the
light-sensing mechanism.
To determine the impact of this variability on the clock precision,
our approach is to investigate how the PRC $V(\phi)$ given by Eq.~\ref{map}
reacts to changes of the daylight driving cycle, as recapitulated in
Fig. \ref{fig1}.
Indeed, changes in amplitude $\epsilon$ and temporal profile $L(u)$
modify the asymptotic entrainment state (Fig. \ref{fig1} {\it A}), and the
resulting variations in phase shift or stability coefficient can be determined
from the associated changes in the PRC  (Fig. \ref{fig1} {\it B}).  

To define relevant measures of robustness, we consider small
changes of daylight intensities with respect to some average value:
\beq
\epsilon L(u) = \epsilon_0 (L_0(u) + \delta \epsilon \tilde{L}(u))
\label{eq:perturbation}
\eeq
where $u\in[0,\tau_D]$, $\tilde{L}$ is an appropriately normalized
perturbation profile and
$\delta\epsilon$ is small compared to 1. Such a perturbation generally
modifies the PRC $V(\phi,\epsilon)$, and consequently the phase shift
$\phi^*$ and the 
stability coefficient $V'(\phi^*)$. Considering that these two
important properties of entrainment should vary
as little as possible,  
two complementary sensitivity measures $\Pi$ and $\Sigma$ can be
defined, which are respectively
the variance of the phase shift and the relative variance of the stability
coefficient in response to small daylight fluctuations of randomly distributed
amplitudes (see Appendix  {\it A}). Assuming that the phase shift and
stability coefficient
depend linearly on amplitude perturbation $\delta\epsilon$,  $\Pi$ and
$\Sigma$ can 
be written:

\begin{equation}
\left\{ \begin{array}{ll}
\Pi &=\,[D_{\delta\epsilon} \phi^*(\epsilon_0)]^2\\
\Sigma &=\, [D_{\delta\epsilon}\partial_{\phi}V(\phi^*,\epsilon_0)/\partial_{\phi}V(\phi^*,\epsilon_0)]^2
\end{array} \right.
\label{sigpi}
\end{equation}

The quantity $\Sigma$ provides indirect information about the
entrainment range of the forcing mechanism.  
Indeed, there are usually minimum and maximum values of the
forcing strength such that
outside the interval they define, complex dynamical regimes such as
quasiperiodic, period-doubled or chaotic ones occur \cite{Pikovsky00}. 
These alternative regimes are associated with a high variability of the phase
shift, which is incompatible with the time-keeping function of the clock.
Achieving stable entrainment for a large spectrum of daylight
intensities therefore requires that the stability coefficient is
sufficiently independent of forcing strength so that it does not approach
 marginal stability ($V'(\phi^*)= 0$ or $-2$) too closely. 
Even when the 1:1 entrainment state is stable over a wide
range, the phase shift $\phi^*$ can still vary in presence of fluctuations of
daylight amplitude and profile, which is measured by $\Pi$.
It is important to emphasize that, because $\Pi$ and $\Sigma$ are defined in
with respect to the asymptotic phase shift, this approach is restricted to the
case where the relaxation time toward the asymptotic state is smaller that the
correlation time of daylight fluctuations over successive days, for instance
when individuals are subjected to slow changes of weather, habitat, 
genotype or phenotype. This approach may nevertheless give reliable
insights into the case where daylight intensities change from day to day.

\subsubsection*{$\Pi$ and $\Sigma$ sensitivities depend on the shape
  of IPRCs}

The sensitivity measures defined by Eq. \ref{sigpi} characterize variations in
phase shift and stability coefficients. These variations depend
themselves on the 
response of the PRC $V(\phi)$ to 
small changes in the amplitude and temporal profile of the forcing
scheme, which must be computed numerically or determined experimentally.
In the limit of weak forcing, however, the PRC can be approximated as a mere 
convolution of $Z$ with the forcing profile (see Eq. \ref{prop}).
Subsequently, $\Sigma$ and $\Pi$ can be expressed in a simple manner
as a function of forcing amplitude $\epsilon$, period mismatch
$\gamma$ and IPRC characteristics.  
In the following, we investigate the robustness with respect to two
types of daylight fluctuations which may occur in nature,
namely changes (i) in the daylight temporal profile or (ii)
in the average daylight amplitude. 

In the first case, the primary effect of the temporal profile changes
in 
light signal is to produce phase-shift variability.
$\Pi$ can then be expressed as the function
of $Z(\phi)$ (see Appendix {\it B}):
\beq
\Pi=\biggl[\,\frac{\int_0^{\tau_D}
Z(u+\phi^*)\,\tilde{L}(u)\,du}{\int_0^{\tau_D}
Z'(u+\phi^*)L_0(u)\,du}\,\biggl]^2
\label{profil}
\eeq
Low values of $\Pi$ are achieved by minimizing the ratio between
the two integrals, 
which constraints function $Z$.
The numerator of $\Pi$ in Eq.~\ref{profil} can be minimized for arbitrary
light input profile fluctuations $\tilde{L}(u)$  when $Z$
is constant over the coupling interval $[\phi^*,\phi^*+\tau_D]$, since
$\int \tilde{L}(u)du=0$.
In contrast to this, maximizing the denominator of $\Pi$ in Eq.~\ref{profil}
typically requires that 
$Z$ decreases significantly over $[\phi^*,\phi^*+\tau_D]$. 
An optimal compromise between these two requirements is reached when
the i-IPRC $Z$ is constant on a large portion of the coupling interval except
at the beginning and the end of the interval, where phase advances or delays
occur, so that the dead zone slightly smaller than the coupling
interval. 
To illustrate this point, Figs. \ref{fig2}A and \ref{fig2}C
show two i-IPRCs computed for two different coupling
schemes of the circadian model
introduced in the following section. The IPRC in Fig.~\ref{fig2} {\it A} 
displays a large dead zone during daytime contrary to the one in
Fig.~\ref{fig2} {\it C}. We will see later that this is associated with
very different degrees of robustness of the two coupling mechanisms. 
Besides $\Pi$, quantities $\Pi_k$ characterizing
phase-shift variance with respect to sinusoidal fluctuations of random
phases can also be defined (see Appendix {\it B}).

Considering now changes in average light intensity, their  effect 
on phase shift variance is easily derived from the property that
$V(\phi^*,\epsilon)=\gamma$ is constant, yielding $\Pi=
[\gamma/V'(\phi^*)]^2$ (see appendix {\it B}).
Entrainment stability is affected in a more subtle manner. We find
that $\Sigma$ can be expressed in terms 
of the
d-IPRC function $W(\phi)$  and its derivatives (see Appendix {\it B}):  
\begin{equation}
  \Sigma= \biggl[\, 1 - \frac{W(\phi^*)
    W''(\phi^*)} {W'(\phi^*)^2} \,\biggl]^2  
  \label{srob}
\end{equation}
where $\phi^*$ satisfies $\epsilon W(\phi^*)=\gamma$.
If the period mismatch $\gamma$ or the second derivative $W''(\phi)$
are zero, $\Sigma=1$. 
Lower (resp., larger) values of $\Sigma$ associated with an extended
(resp., reduced) stability domain of the entrainment state imply
$\gamma W''(\phi^*)>0$ (resp., $<0$). Thus, robustness requires a convex 
PRC when FRP is larger than the forcing period, and a concave one in the
opposite case.
Again, Figs.~\ref{fig2} {\it B} and {\it D} show two d-IPRCs computed
from the circadian model introduced in next section.
The fact that their curvatures are opposite near CT0 suggests that they have
different robustness properties, as we will see later.

In this section, we have derived and discussed two simplified expressions for
$\Pi$ and $\Sigma$ which are valid for fluctuations in daylight
temporal profile and in average daylight intensities, respectively.
This is because changes in the daylight temporal profile, unlike those in the
average daylight intensities, tend to be uncorrelated over successive days,
thus perturbing the phase-shift without destabilizing the
entrainment state. For fluctuations in average daylight intensities,
which may persist over several days or even be permanent
(corresponding to differences between two individuals),
destabilization is the most disrupting effect.
In following sections, computation of $\Pi$ (resp., $\Sigma$) will
thus be
carried using 
Eq. \ref{profil} (resp., Eq. \ref{srob}) to characterize robustness to
fluctuations in daylight temporal
profile (resp., intensity).

\subsubsection*{Robustness to daylight fluctuations in a minimal
  model of circadian oscillator} 

The criteria derived above to  ensure  robust entrainment in the face
of daylight fluctuations hold for any nonlinear oscillator subjected to weak
enough forcing.
In this section, we check whether these criteria also apply to a minimal
circadian oscillator model where coupling to light is generally
non-negligible, and for which the weak forcing approximation does not necessarily
hold.
In most organisms, endogenous clock oscillations appear to rely on
a core negative feedback loop, through which a clock gene encodes proteins
that activate (vs inactivate) its own transcriptional repressor (vs activator)
\cite{Young01}. 
The presence of delays or nonlinearities along the loop favors the emergence
of oscillations in this autoregulatory loop \cite{Novak08,Morant09}.  
This basic clock architecture can be captured in low-dimensional dynamical models
such as the one originally proposed by Leloup and Goldbeter 
for Neurospora clock \cite{Leloup99}: 
a gene sequence is transcribed into mRNA ($M$), which translates into a protein
in the cytoplasm ($P_c$), further
translocated in the nucleus ($P_N$) where it inactivates the gene:

\begin{equation}
\left\{ \begin{array}{rll}
\tau \frac{dM}{dt}   & =  s_M \frac{K_I^n}{K_I^n+P_N^n} - d_M
\frac{M}{K_M+M}\\  
\tau \frac{dP_C}{dt} & =  s_P M - d_P \frac{P_C}{K_P+P_C}-k_1\,P_C+k_2\,P_N \\
\tau \frac{dP_N}{dt} & =  k_1 P_C - k_2 P_N
\end{array} \right.
\label{neuro}
\end{equation}

The Michaelis-Menten-like kinetics used to describe transcription and
degradation dynamics is required for the appearance of spontaneous
oscillations. 
We use the following model parameters, which give rise to $24$-hour
oscillations in the dark: $n=4$, $s_M=2.2$, $K_I=1.8$, $d_M=0.84$, $K_M=0.5$,
$s_P=0.4$, $d_P=1.6$, $k_P=0.13$, $k_1=0.4$, $k_2=0.45$, $\tau=1$.
The effect of light is to modify one or several parameters
during a time interval $\tau_D$ and with gain $\epsilon$ as in
Eq. \ref{forcing}.
To investigate the influence of a mismatch between the forcing period
$T$ and the FRP $T_{0}$, the latter is varied by
changing the time constant $\tau$ ($T_{0}=24\,\tau$).

To illustrate our analysis, we have selected two
light-coupling schemes  on the basis of their IPRCs so that they provide us
with examples of robust and non 
robust entrainment 
with respect to daylight fluctuations.
These two schemes consist in the activation or repression of transcription
by light, corresponding respectively to an increase or a decrease of $s_M$ during daytime.
The IPRCs associated with transcriptional
repression by light indicates an insensitivity during the time
interval in which 
light is applied (Fig. \ref{fig2} {\it A}) and a negative sign of the third
derivative  near its zero (Fig. \ref{fig2} {\it B}). According to Eqs.
\ref{profil} and \ref{srob}, both properties are 
expected to favor 
entrainment robustness. 
The opposite conclusions hold for the IPRCs associated with
transcriptional activation by light (Figs. \ref{fig2} {\it C}, {\it D}), which 
presumably indicates poor robustness properties.

Fig. \ref{fig3} confirms that these two forcing schemes significantly 
differ in the robustness of the entrainment state to variations in
light intensity, as we
anticipated from their IPRCs. 
In case of transcriptional repression by light, entrainment remains
stable at various light levels (left panel of
Fig. \ref{fig3} {\it A}), and the entrained oscillations vary little.  
In contrast to this, period doubling occurs when transcription is
activated by light (right panel of Fig. \ref{fig3} {\it A}).

As predicted by the theory, these different robustness properties of the two
forcing schemes when light level changes for a given FRP can be traced
back to how the PRC and its derivative change when light levels
are increased (Fig. \ref{fig3} {\it B}). In the case of transcriptional
repression by light, the PRC varies very little in an extended
neighborhood of the fixed point.
In particular, $V'(\phi^*)$  remains well within the stability boundary,
whereas in the case of transcriptional activation by light, the
phase-locked state is destabilized 
($V'(\phi^*)<-2$) beyond a critical forcing amplitude where period-doubling
typically occurs (Fig. \ref{fig3} {\it C}).
Accordingly, values of $\Sigma$ are much lower in the case of
transcriptional repression by light at all forcing strengths
(Fig. \ref{fig3} {\it D}). 
Fig. \ref{fig3} {\it E} shows that these robustness properties do not depend
on the FRP. Whatever the value of the latter,
the 1:1 entrainment domain is much wider for transcriptional
repression by light. 
In fact, the main effect of the period mismatch is to force
a minimum modulation amplitude for synchronization when it it
increases or decreases away from 0. 

These two light-coupling schemes are also associated with different robustness
properties of the clock phase with respect to changes of the daylight temporal
profile (Fig. \ref{fig4}).  
The phase shift variance induced by small sinusoid  
perturbations of different phases is much smaller (by a factor from 2 to
10) when light represses rather than activates the self-regulated
gene (Fig. \ref{fig4} {\it A}).
Again, this difference in robustness can be explained by the
sentitivity or insensitivity of the PRC near the stationary phase
shift (Fig. \ref{fig4} {\it B}), 
which ultimately depends on the existence of dead zone in the relevant
time interval (Fig. \ref{fig2} {\it A} and {\it C}).
Fig. \ref{fig4} {\it C} shows that these differences in phase-shift
sensitivities with respect to sinusoidal profile fluctuations are more or less
pronounced  depending on the specific value of the FRP $T_0$, or of the
characteristic period of the fluctuations ($k=1$ or $2$). 

Taken together, Figs. \ref{fig3} and \ref{fig4} shows two light-coupling
schemes that are associated with a large (resp. small) entrainment 
domain inside which there is a low (resp. large) phase-shift variability.
In fact, the two quantities tend to be correlated, as is assessed by a
systematic correlation analysis (see Fig. S1 in the Supporting Material).

\subsubsection*{Relating robustness to IPRC: validity of the weak coupling
  approximation} 

The extent to which IPRC properties can account for the robustness of
circadian clock with respect to daylight fluctuations depends on the
validity of the phase approximation.  
In this section, we therefore evaluate the agreement between robustness
properties derived analytically in the limit of weak forcing and those
measured numerically when the forcing is not weak, as depicted in Fig. 
\ref{fig5}.

A first indication is provided by estimating  the light-stimulus
amplitude beyond which the proportional relationship between phase
response and light 
amplitude (Eq. \ref{prop}) no longer holds. 
In this respect, we found that the phase approximation can remain
acceptable for 
modulations in excess 50\%, depending on the parameter
considered and on the stimulus phase (Fig. \ref{fig5} {\it A}, {\it B}).
Other quantities derived from PRCs, like the phase
shift $\phi^*$, the stability coefficient $V'(\phi^*)$, or the
phase-shift and coefficient stability sensitivities $\Pi$ and $\Sigma$, are
also expected to deviate from their estimates in the phase approximation.
Estimation of the phase shift and the stability coefficient of the
entrainment state is quantitatively good when parameters change by 30\%
between night and day (Fig. \ref{fig5} {\it C}, {\it D}). 
Quantitative agreement is slightly more difficult for $\Pi$
and $\Sigma$ (Fig. \ref{fig5} {\it E}, {\it F}), but a qualitative agreement
can still be observed and is in fact more than sufficient to reflect a large
(over 20-fold) dispersion of values of $\Pi$ and $\Sigma$ between the most 
and least robust entrainment schemes. 

The study of the range of validity of our approach has been extended
to more realistic circadian clock models associated with various 
organisms like Drosophila \cite{Leloup99}, Ostreococcus \cite{Thommen10},
cyanobacteria \cite{Rust07} or mammals \cite{Geier05} (see Fig. S2 in the
Supporting Material).  
The good agreement  observed between robustness measures and their
estimates using IPRC confirms that the phase approximation
remains qualitatively valid for moderate forcing strength regardless
of the model 
properties, thus providing an efficient tool to discriminate forcing schemes
according to their robustness properties.

\subsubsection*{Experimental PRCs analysis reveal robustness properties to
  daylight fluctuations} 

We have shown that PRCs, despite of their simplicity, can provide a
detailed 
and reliable information on the robustness properties of circadian
entrainment.
Incidentally, PRCs have been measured over fifty years in many organisms
\cite{Johnson90}, thereby offering indirect evidence of whether
natural clocks 
are robust or not to daylight fluctuations.

We have selected several PRCs measured in response to light pulses of various
durations for twelve species
\cite{Konopka79,Honma85,Daan76,Perlman81,Covington01,Fuentes83,Eskin71,Christ73,Johnson89,Saunders78,Saunders77} 
(Fig. \ref{fig6} {\it A}). 
One may assume that for type-I experimental PRCs (a to j panels), the phase
approximation holds. 
Therefore, their corresponding impulse IPRCs are expected to display a shape
similar to that of experimental PRCs if the light pulse is short enough
(panels {\it a}-{\it g} of Fig. \ref{fig6} {\it A}) or can otherwise be
obtained by a rectangular-pulse 
deconvolution (panels {\it h}-{\it j} of Fig. \ref{fig6} {\it A}).
In any case, the impulse IPRCs display pronounced dead zones
(Fig. \ref{fig6} {\it B}), albeit of different sizes, which is likely to confer
robustness to daylight fluctuations according to our results.

For a quantitative assessment of this conjecture, we compare the
values of $\Pi$ for the i-IPRCs estimated from experimental
data with those obtained for many different light-coupling schemes in
various circadian models (Fig. \ref{fig6} {\it C}).
Values of $\Pi$ obtained from experimental PRCs are found to be
systematically lower than in the
control case of a linear i-PRC, and are consistent with those computed for
the most robust entrainment schemes of various circadian models.
This
strongly supports the idea that the circadian clock of these organisms have
evolved robustness properties with respect to daylight fluctuations.

PRCs in response to daylight-like stimuli such as $12$-hour rectangular pulse
also provide valuable information about the entrainment robustness measured
by the quantity $\Sigma$. 
Unfortunately, those PRCs have been much less measured experimentally, and can
be hardly estimated from the PRCs available without a precise knowledge of the
light-coupling profile.
Noticeable exceptions are the PRCs measured in two species
\cite{Saunders77,Saunders78} that are depicted in the panels (k) and (l) of
Fig. \ref{fig6} {\it A}. 
The apparent change of curvature from convex to concave when the circadian
phase goes from negative to positive seems again to indicate
a robust entrainment.
 
\section*{DISCUSSION}%

In most living organisms, a circadian clock synchronizes
internal physiological processes with cyclic environmental changes
(e.g, in light
or temperature) associated with Earth rotation. 
A stable phase relationship between internal and external time is however
challenged by cellular and environmental variability. Part of this
variability is linked to the diurnal cycle, in particular  fluctuations in
daylight perceived. 
In this work, we have derived explicit criteria for the robustness of
circadian entrainment to daylight variability.  
Combining analytical and numerical approaches, we have shown that 
robustness depends critically on the precise shape of the IPRC, which 
characterizes the linear response of an oscillator to vanishingly small light
inputs.
On the one hand, the existence of an extensive range of entrainment is favored
by a convex (resp. concave) d-IPRC (IPRC for daily pattern of light) when the
FRP is larger (resp. lower) than the day-night driving period.  
On the other hand, a low dispersion of the phase shift between the internal
oscillator and the environmental driving cycle requires a phase
response insensitivity of the clock during daytime, the so-called dead zone,
which is manifested by the null interval of the i-IPRC (IPRC for light
impulse) during the subjective day.

These two PRC properties confering entrainment robustness
to forcing fluctuations happen incidentally to be observed in circadian
PRCs measured experimentally in many organisms, in particular the
existence of a dead zone that seems to be universal.
Although clock resetting toward a stable phase shift obviously
requires that phase advances and delays in PRCs occur in early and late
day respectively, the presence of a dead zone has apparently emerged as a
strategy to avoid phase variability due to daylight fluctuations and to ensure
entrainment stability. 
The different size of dead zone intervals observed in various species does not
weaken the robustness hypothesis but rather suggests the existence
of  light-gating mechanisms that restrict the light sensitivity of the clock
to a certain window of the day \cite{Geier05,Thommen10}.
These universal PRC properties of circadian clock may reflect an universal
strategy to minimize the impact of fluctuations in the forcing cycle on the
clock phase, thus suggesting a convergent functional evolution of circadian
clocks.

\subsubsection*{Trade-offs between multiple evolutionary goals}

Circadian clock properties, including IPRC or FRP properties, are also
constrained by other adaptive purposes besides the need for robustness to
daylight fluctuations, raising the question of whether trade-offs are
eventually required.

One important evolutive constraint for circadian clock is the requirement to
achieve fast resetting after jetlag \cite{Granada09a}, which depends in
first approximation on the derivative of the PRC at $\phi^*$ (optimal
resetting is obtained for $V'(\phi^*)=-1$).  
For circadian oscillators that display a dead zone, the weak forcing
approximation indicates that fast resetting requires that
the dead zone must be slightly smaller than the daylight interval
or/and that the FRP must 
be different from 24 hours. These two conditions are fully compatible with
criteria for robustness to daylight fluctuations.

The variability of the FRP of the circadian rhythm is also a prevalent
source of variability susceptible to affect clock phase \cite{Daan76}.
Daan and Pittendrigh stressed the fact that a clock characterized by
(i) a FRP equal to 24 hours and (ii) a large dead zone of the PRC 
displays a phase instability for which a small change in the FRP causes a
large change in phase shift.
This result has been tempered by a modeling study \cite{Kurosawa06} showing
that entrained circadian oscillations for the Drosophila model, which also
display a dead zone, have a wider range of stability when the FRP is close to
$24$ hours.
These apparently contradictory results can be reconciled in the light of our
results when reasoning in terms of d-IPRCs of the circadian oscillator and
depending on whether one focus on the effect of the FRP variability on the
destabilization of the 1:1 synchronization state or on the phase-shift
dispersion.
The argument of Daan and Pittendrigh relies on the assumption that the first
derivative of the d-IPRCs is small (i.e., the dead zone of the i-IPRC is larger
than the light-coupling interval) whereas the argument of Kurosawa and Golbeter
depends primarily on the second derivative values of the d-IPRC.
In fact, it is likely that any circadian oscillators is characterized by an
optimal value of the FRP whose distance from 24 hours would achieve a
compromise in maintaining phase-shift stability and precision in face of both
sources of variability in daylight intensities and FRP. 

Finally, seasonal variations of daylength is another environmental
variation that the clock needs to adapt to by tracking dawn, dusk or
both. 
Previous studies \cite{Geier05,Gunawan07} have shown that the
presence of a dead zone is an efficient mechanism to achieve seasonal
tracking of dawn (resp. dusk) with a FRP larger (resp. smaller) than 24 hours.     
However, this mechanism requires that the dead zone is larger than the light
coupling interval for some short photoperiods, which may antagonize the
requirement for a robust synchronization in presence of daylight
fluctuations.

Thus, circadian clocks characterized by a dead zone during the subjective day
and a finely tuned FRP, eventually supplemented with specific
light-gating mechanisms \cite{Geier05,Thommen10}, are suitable to achieve an
efficient trade-off between multiple and unrelated evolutionary goals.

\subsubsection*{Toward design principles of circadian clocks}

The concept of phase response curve was introduced and exploited
to investigate theoretically and experimentally rhythmic and synchronization
behaviors in various biological systems
\cite{Winfree01,Granada09b}. 
When oscillators are weakly perturbed, their PRC reduces to an infinitesimal
analog whose linearity allows a more comprehensive analysis of the
synchronization behavior \cite{Kuramoto84,Rand04,Taylor08}. 
Applying this approach in the context of circadian clocks allowed us to
characterize the robustness of circadian entrainment with respect to
daylight fluctuations in terms of specific characteristics of infinitesimal
PRCs. 
This finding opens up a broad range of issues regarding the biophysical
implementation of specific phase response curve properties in circadian
clocks, similarly to the role of ionic currents in shaping the phase response
curve of neuronal oscillators  \cite{Pfeuty03,Acker03}.
So far, many features of oscillatory behavior have been related
to the structural basis of the underlying biochemical oscillator
\cite{Wang08,Novak08,Hasty08,Tsai08,Pfeuty09}, but not to phase response curve
characteristics.
Understanding how the phase response curve is shaped by various
features of light coupling, including gating 
\cite{Geier05} and photo-adaptation \cite{Malzahn10}, or intrinsic oscillator
properties such as feedback, saturation or delays, would undoubtely
shed light on the role of some key 
features of circadian clock designs across species.

\section*{Acknowledgment}%

The authors acknowledge support of from Agence Nationale de la
Recherche (ANR) under reference 07BSYS004 and from CNRS through the
Interdisciplinary Programme ``Interface Physique-Chimie-Biologie: aide
\`a la prise de risque''.

\newpage

\section*{Appendices}%

\subsection*{Appendix A. General expression of robustness quantities} 

The first-return map in Eq.~\ref{map} approximates the dynamics of a 
periodically forced non-linear oscillator and introduces the function
$V(\phi)$, namely a PRC,  which is an implicit function of amplitude
$\epsilon$ and temporal profile $L$ of the light forcing.
We consider first that forcing properties slightly vary among individuals
with respect to some average daylight forcing $\epsilon_0 L_0(u)$:
\beq
\epsilon L(u)=\epsilon_0 (L_0(u) + \delta \epsilon \,\tilde{L}(u))
\eeq
where $L$ (or $L_0$) is normalized with $1/\tau_D \int_0^{\tau_D}
L(t)dt=1$. This normalization must be preserved by an appropriate normalization
of $\tilde{L}$ that depends on the type of fluctuations
considered (appendix {\it B}). 

We can now expand the phase shift $\phi^*$ and the stability coefficient
$\partial_{\phi}V(\phi^*)$ up to first order in $\delta \epsilon$ :
\beq
\left\{ \begin{array}{ll}
\phi^* (\epsilon_0,\delta\epsilon) &=\, \phi^*(\epsilon_0,0) + \delta\epsilon
    [D_{\delta\epsilon} \phi^*(\epsilon_0,0)] 
+ 0(\delta\epsilon^2)\\
\partial_{\phi} V(\phi^*,\epsilon_0,\delta\epsilon) &=\, 
\partial_{\phi} V(\phi^*,\epsilon_0,0)
 + \delta\epsilon
 [D_{\delta\epsilon}\partial_{\phi}V(\phi,\epsilon_0,0)] 
 + 0(\delta\epsilon^2)
\end{array} \right.
\label{sigpi0}
\eeq
We introduce the quantities $\Pi$ and $\Sigma$ defined as the relative 
variances of the phase shift and the stability coefficient induced by daylight 
fluctuation of random amplitude:
\begin{equation}
\left\{ \begin{array}{ll}
\Pi
&=\,\frac{<(\phi^*(\epsilon_0,\delta\epsilon)-\phi^*(\epsilon_0,0))^2>}{<\delta \epsilon^2>} \\    
\Sigma &=\,\frac{<(\partial_{\phi}
V(\phi^*,\epsilon_0,\delta\epsilon)-\partial_{\phi} V(\phi^*,\epsilon_0,0))^2>}{<\delta
  \epsilon^2><\partial_{\phi} V(\phi^*,\epsilon_0,0)^2>}  
\end{array} \right.
\label{sigpi1}
\end{equation}
The approximation of small fluctuation amplitudes in Eq. \ref{sigpi0} allows
to rewrite those quantities as following:
\begin{equation}
\left\{ \begin{array}{ll}
\Pi &=\,[D_{\delta\epsilon} \phi^*(\epsilon_0,0)]^2\\
\Sigma &=\, [D_{\delta\epsilon}\partial_{\phi}
  V(\phi^*,\epsilon_0,0)/\partial_{\phi}
  V(\phi^*,\epsilon_0,0)]^2
\end{array} \right.
\label{sigpi2}
\end{equation}
According to the fixed point condition of the map (Eq. \ref{map}),
$V(\phi^*(\delta\epsilon),\delta\epsilon)=V(\phi^*(0),0)=\gamma$.
Expanding $V(\phi^*,\delta\epsilon)$ up to first order in $\delta\epsilon$
leads to: 
\beq
\partial_{\delta\epsilon}V(\phi^*,\epsilon_0,0) +
V(\phi^*,\epsilon_0,0)\partial_{\delta\epsilon} 
\phi(\epsilon_0,0)=0
\label{chain}
\eeq
$\Pi$ can be therefore expressed uniquely as a function of the PRC, $V(\phi)$:      
\beq
\Pi =
\biggl[\,\frac{\partial_{\delta\epsilon}
    V(\phi^*,\epsilon_0,0)}{\partial_{\phi}  
 V(\phi^*,\epsilon_0,0)}\biggl]^2
\label{sigpi3}
\eeq
Combining Eqs. \ref{sigpi2} and \ref{chain} allows to develop the
quantity $\Sigma$ as follows: 
\beq
\Sigma = \biggl[ \,\frac{\partial_{\phi} \partial_{\delta\epsilon}
    V(\phi^*,\epsilon_0,0)}{\partial_{\phi} V(\phi^*,\epsilon_0,0)}
 -  \frac{\partial_{\delta\epsilon} V(\phi^*,\epsilon_0,0)\partial_{\phi}^2
  V(\phi^*,\epsilon_0,0)}{[\partial_{\phi} 
 V(\phi^*,\epsilon_0,0)]^2} \,\biggl]^2
\label{sigpi4}
\eeq

\subsection*{Appendix B. Approximation of robustness quantities in the weak
  forcing limit} 

Assuming that the forced periodic orbit remains in a close
neighbourhood of the free-running periodic orbit, we can use the phase
approximation, 
according to which:
\beq
V(\phi,\epsilon_0,\delta \epsilon)=\epsilon\,[W(\phi)+\delta\epsilon \,\tilde{W}(\phi)]
\label{w_app0}
\eeq 
where 
\beq
W(\phi)=\int_0^{\tau_D} L(u) Z(u+\phi)du 
\label{w_app1}
\eeq
and
\beq
\tilde{W}(\phi)=\int_0^{\tau_D} \tilde{L}(u) Z(u+\phi)du 
\label{w_app2}
\eeq

Substituting Eq. \ref{w_app0} in Eqs. \ref{sigpi3} and \ref{sigpi4}
allows one to express the measures of robustness introduced above  as
functions of $W$ and $\tilde{W}$: 

\beq
\Pi = \biggl[\,\frac{\tilde{W}(\phi^*)}{W'(\phi^*)}\,\biggl]^2
\label{eq14}
\end{equation}
and
\begin{equation}
\Sigma = \biggl [1\,-\, \frac{\tilde{W}(\phi^*)
    W''(\phi^*)}{W'(\phi^*)^2} \,\biggl]^2
\label{eq15} 
\eeq

\subsubsection*{a. Fluctuations in the average daylight intensities}

The simplest example of fluctuations in the light driving cycle corresponds to
the case where the overall gain of the light signal fluctuates.
The normalization requirement for $\tilde{L}(u)$ imposes that $\tilde{L}(u) =
L(u)$ and $\tilde{W}(\phi)=W(\phi)$, which allows to rewrite the sensitivity
quantities of Eqs. \ref{eq14} and \ref{eq15} 
as: 
\begin{equation}
\Pi= \biggl[\,\frac{\gamma}{\epsilon W'(\phi^*)}\,\biggl]^2
\end{equation}
and
\begin{equation}
\Sigma = \biggl[1-\, \frac{\gamma W''(\phi^*)}{\epsilon
  W'(\phi^*)^2} \,\biggl]^2
\end{equation}

\subsubsection*{b. Fluctuations in daylight profiles}

Alternatively, one can also consider the case where only the temporal profile
change with some normalized variance whereas the light intensity averaged over
the day remains unchanged: 
\beq
1/\tau_D \int_0^{\tau_D}\tilde{L}(u)^2\,du=1
\eeq 
and
\beq
\int_0^{\tau_D}\tilde{L}(u)\,du=0
\eeq 
Note that this normalization of $\tilde{L}$ preserves the normalization of
$L(u)$. 
Substituting Eqs. \ref{w_app1} and \ref{w_app2} in Eq. \ref{eq14}
allows us to
rewrite $\Pi$ as a function of $Z$ instead of $W$: 
\beq
\Pi=\biggl[\,\frac{\int_0^{\tau_D}
Z(u+\phi^*)\,\tilde{L}(u)\,du}{\int_0^{\tau_D}
Z'(u+\phi^*)L(u)\,du}\,\biggl]^2
\label{eq17}
\eeq
One can also estimate the phase-shift sensitivity in response to sinusoidal 
daylight fluctuations of period $\tau_D/k$ and phase $\psi_k$. 
Decomposing $\tilde{L}(u)$ as a Fourier series:
\beq
\tilde{L}(u,\psi)=\sum_k \, \tilde{l}_k cos(u k/\tau_D +\psi_k)
\label{Fourier}
\eeq 
and substituting Eq. \ref{Fourier} in Eq. \ref{eq17} leads to:
\beq
\Pi(\psi)=\biggl[\frac{\sum_k a_k\,cos(\psi_k)+b_k\, sin(\psi_k)}{\int_0^{\tau_D}
Z'(u+\phi^*)L(u)\,du}\,\biggl]^2
\eeq
where $a_k$ and $b_k$ as the $k$th cosine and sine Fourier coefficients
of the i-IPRC truncated on the subinterval in which daylight perturbs
the clock (usually daytime).
Summing over $\psi_k$ between $0$ and $2\pi$ gives the averaged phase shift
associated with a arbitrary fluctuations of zero mean ($\delta l_0=0$) and
unitary norm ($\sum_k \tilde{l}_k^2 = 1$):
\begin{equation}
\Pi=<\Pi(\psi)>_{\psi}= 1/2\pi\int_0^{2\pi}
\Pi(\psi)d\psi=\sum_k \tilde{l}_k^2 \,\,\Pi_k  
\label{app_gen}
\eeq
where $\Pi_k$ are the phase-shift variances associated with sinusoidal
fluctuations of period $\tau_D/k$ :
\beq
\Pi_k=\frac{a_k^2\,+b_k^2}{2\,W'(\phi^*)^2}
\label{app_pi}
\eeq
These quantities are for instance easily computed in the case where $Z$ is a
decreasing linear function on the interval of coupling, which leads to
$\Pi_k=\tau_D^2/8\,k^2\,\pi^2$ for $L(u)=1$ during daytime.
          

\subsection*{Appendix C. Robustness analysis of experimental PRCs} 

In this section, we briefly describe the procedure used to analyse the 
experimental PRCs.
First, the fitting procedure adjusts the discrete experimental data
$t_j,y_j$ with $j=1,N$ by a continuous function $f(t)$. 
It is based on the minimization of both the fitting error and the second
derivative of $f$:
\beq
S_1= \frac{k_1}{N} \sum_{j=1,N} (y_j-f(x_j))^2 + \frac{k_2}{T} \int_0^T
[f''(t)]^2\,dt  
\eeq
The ratio $k_2/k_1$ is adjusted typically between $5$ and $15$ according to
the data. 

For experimental PRCs that are measured using relatively short light pulse of
less than one hour, the estimated i-IPRC, $z$, is assumed to be roughly equal
to $f$.
Otherwise we perform a deconvolution operation to extract the estimated
i-IPRCs, using a genetic algorithm to find the function, $z$, that minimize
the error quantity:
\beq
S_2=\int_0^{T} \biggl[f(t) - \int_t^{t+\tau_D} z(u)du \biggl]^2 dt
\eeq
To estimate $\Pi$-values associated with experimental PRCs, we use
Eq. \ref{app_pi} with the estimated i-IPRCs $z$.

\newpage

\section*{Figure Legends}

\subsubsection*{Figure~\ref{fig1}.}
{\bf Measures of clock robustness to daylight fluctuations}.
    ({\it A}) Example of two distinct daylight profiles ({\it top panel})
    leading to different phase shifts for the entrained oscillators ({\it
    bottom panel}). 
    The shaded area corresponds to night. 
    ({\it B}) In the top panel, plots of the PRCs $V(\phi)$ measuring the phase
    change induced by the daylight perturbations shown in A applied at
    different phases.
    The stable phase shifts $\phi^*$ in the entrainment regime are
    solutions of $V(\phi)=\gamma$ defining . 
    Infinitesimal values of $\Delta \phi^*$ with respect to small forcing 
    discrepancies defines the sensitivity measure $\Pi$.
    In the bottom panel, plot of the derivative of the two PRCs, from
    which the stability coefficients $V'(\phi^*)$ of the entrainment states
    and their difference can be inferred. 
    The shaded area corresponds to stable entrainment, associated with
    $-2<V'(\phi^*)<0$.
    Infinitesimal values of $\Delta V'(\phi^*)$ with respect to small
    forcing discrepancies defines the sensitivity measure $\Sigma$.

\subsubsection*{Figure~\ref{fig2}.}
{\bf Two IPRCs associated with different 
    robustness properties and their characteristics.}  
    Plots of IPRCs in a circadian model where light induces either a
    repression ({\it A},{\it B}) or an activation ({\it C},{\it D}) 
    of clock gene transcription.
    In these examples, FRP is $24$ hours ($\gamma=0$).
    ({\it A},{\it C}) Plot of i-IPRCs $Z(\phi)$  measured in response to light
    impulses. Whether or not they display a dead zone during the
    light-coupling interval ({\it full line}) determines the
    phase-shift sensitivity $\Pi$ to daylight fluctuations according to the
    Eq. \ref{profil}. 
    ({\it B},{\it D}) Plots of d-IPRCs $W(\phi)$ measured in response to 
    $12$-hour rectangular light pulses. 
    Slight changes of $\gamma$ modify the entrainment phase shift $\phi^*$ 
    (that would correspond to a displacement along the {\it full line} of
    $W$). 
    Whether or not the $W''(\phi^*)>0$ (resp., $<0$) ({\it inset panel})
    when $W(\phi^*)>0$ (resp., $<0$) determines the sensitivity $\Sigma$ of the
    entrainment stability to daylight fluctuations according to the
    Eq. \ref{srob}.

\subsubsection*{Figure~\ref{fig3}.}
{\bf Range of stable 1:1 entrainment in the face of   light-amplitude 
  fluctuations} 
    ({\it A}) Time course over one day of the $M$ clock component ({\it full
    lines}) for different light forcing amplitudes
    (green: $\epsilon=0.3$; red: $\epsilon=0.6$;
    blue: $\epsilon=0.9$) 
    which modulate negatively ({\it left panel}) or positively ({\it right
    panel}) $s_M$ ({\it dashed line}). 
    The shaded area corresponds to night.
    The FRP is set to $T_0=25$ hours.
    ({\it B}) Plots of $V(\phi)$ and $V'(\phi)$ corresponding to the
    different forcing  
    strengths shown in ({\it A}) ({\it same color code}).
    The shaded area corresponds to a stable 1:1 entrainment state.
    ({\it C},{\it D}) Plots of $V'(\phi^*)$ and $\Sigma$ as a function of
    $\epsilon$ for $T_{0}=25$ hours.  
    ({\it E}) Phase diagram showing the different dynamical regimes that can
    be observed as a function of the forcing amplitude $\epsilon$ and
    period mismatch $\gamma$,
    extrapolated from the existence of a stable phase shift $\phi^*$
    and the value of $V'(\phi^*)$ (1:1 Sync: 1:1 entrainment,
    $-2<V'(\phi^*)<0$ ; QP : quasi-periodicity,
    $V(\phi^*)=\gamma$ has  no solution ; CS : complex synchronization and
    chaos, $V'(\phi^*)<-2$). 

\subsubsection*{Figure~\ref{fig4}.}

{\bf Sensitivity of phase-shift to light-profile fluctuations.} 
    Left and right panels correspond to negative and positive
    modulation of $s_M$  by light during daytime with $\epsilon=0.3$.
    ({\it A}) Time courses of the light-dependent parameter and of the
    entrained circadian 
    oscillations over one day in presence of sinusoidal daylight fluctuations. 
    The FRP $T_0$ is set to $24$ hours.
    ({\it B}) Changes in PRCs associated wit fluctuations shown in A. 
    The thick red lines indicate the range of phase shifts ($\gamma=0$).
    ({\it C}) Corresponding values of $\Pi_k^{1/2}$ as a function of $\gamma$ 
    ({\it full line}: $k=1$; {\it dashed line}: $k=2$).

\subsubsection*{Figure~\ref{fig5}.}

{\bf Validity of the weak forcing approximation.}
    ({\it A}, {\it B}) Plot of normalized PRCs $V(\phi)/\epsilon$ for various
    forcing strengths ranging from $\epsilon=0.05$ ({\it black line}) to
    $\epsilon=0.5$ ({\it red line}).
    In the presence of light, $s_M$ decreases in A and increases in B. 
    ({\it C}-{\it F}) Plots of various measures of synchronization and their
    estimates using 
    IPRCs computed for all possible one-parameter modulations, with a
    light-modulation strength 
    of $\epsilon=0.3$ and FRP of $T_{0}=23.7$ and $24.3$ hours. 
    ({\it C}) Phase shifts $\phi^*$.
    ({\it D}) Stability coefficient $V'(\phi^*)$.
    ({\it E},{\it F}) Square roots of $\Sigma$ and $\Pi$. Dashed lines
    correspond to the control case of an IPRC linearly decreasing during the
    daylight interval.

\subsubsection*{Figure~\ref{fig6}.}

{\bf Experimental PRCs reveal robustness properties.} 
    ({\it A}) Experimental PRC data ({\it triangles}) collected from many
    organisms \cite{Johnson90} and fitted ({\it full line} and {\it
    shaded area}) according to the procedure described in the Appendix {\it
    C}. 
    ({\it a}) Drosophila melanogaster, $40$ min pulse \cite{Konopka79}; 
    ({\it b}) Rattus albicus, $30$ min pulse \cite{Honma85};
    ({\it c}) Mus musculus, $15$ min pulse \cite{Daan76}; 
    ({\it d}) Mesocricetus Auratus, $15$ min pulse \cite{Daan76}; 
    ({\it e}) Neuropora crassa, $5$ min pulse \cite{Perlman81};
    ({\it f}) Arabidopsis Thaliana, $1$ hour pulse \cite{Covington01}; 
    ({\it g}) Procambarus bouvieri, $15$ min pulse \cite{Fuentes83}; 
    ({\it h}) Passer domesticus, $6$ hour pulse \cite{Eskin71};
    ({\it i}) Gonyaulax polyedra, $3$ hour pulse \cite{Christ73};
    ({\it j}) Paramecium Bursaria, $4$ hour pulse \cite{Johnson89}
    ({\it k}) Nauphoeta cinerea, 12 hours pulse \cite{Saunders77};
    ({\it l}) Sarcophaga argyrostoma, 12 hour pulse \cite{Saunders78}. 
    ({\it B}) Estimated i-IPRCs associated from the fitted data of the panel
    ({\it a}) to ({\it j}) of ({\it A}).
    ({\it C})
    Plots of the ranges of $\Pi_{k=1}^{1/2}$-values ({\it a-j}, {\it shaded
    area}) computed from estimated i-IPRCs shown in ({\it B}) related with
    data shown 
    in ({\it A}) 
    associated with some range of FRP around 24 hours. For
    comparison, plots of the ranges of $\Pi_{k=1}^{1/2}$-values ({\it k},
    {\it shaded area}) computed for  all coupling schemes of the computational
    circadian clock model (see Fig. \ref{fig5} {\it F}). 
    The horizontal solid line indicates the value obtained for a linear IPRC
    (i.e., with zero curvature).

\clearpage

\begin{figure}  
  \centering \includegraphics*[width=4in]{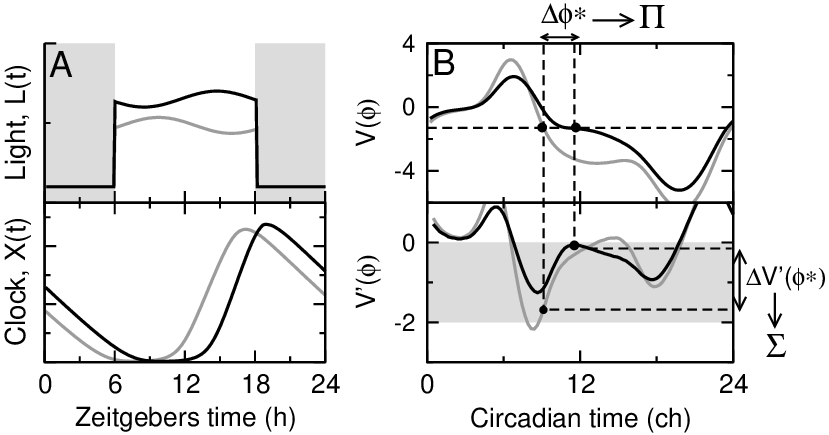} 
  \caption[]{} 
  \label{fig1}
\end{figure}

\clearpage

\begin{figure}  
  \centering
  \includegraphics*[width=3.25in]{fig2.eps} 
  \caption[]{} 
  \label{fig2}
\end{figure}

\clearpage

\begin{figure}  
  \centering \includegraphics*[width=5in]{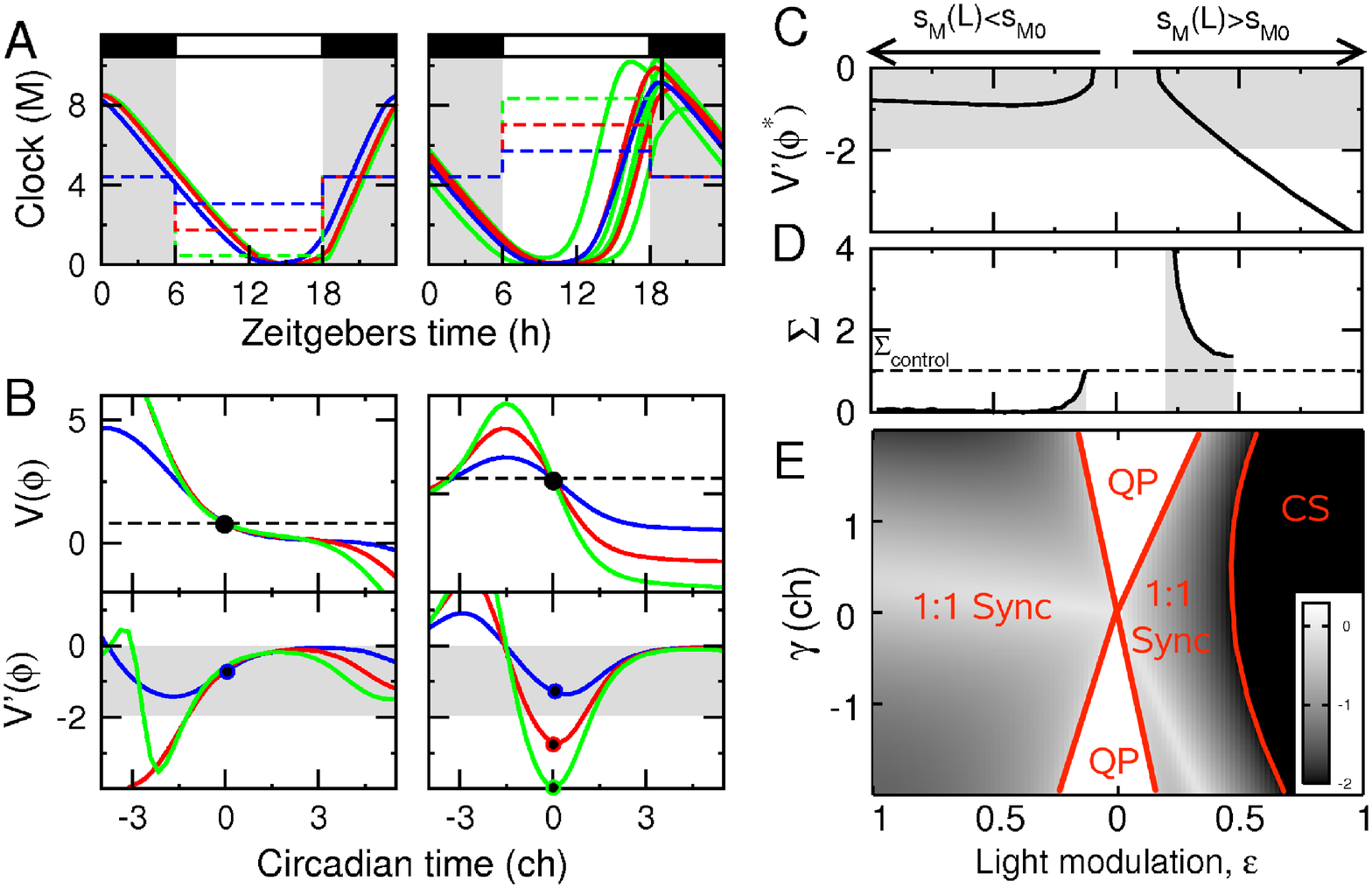} 
  \caption[]{}
  \label{fig3}
\end{figure}

\clearpage

\begin{figure}  
  \centering \includegraphics*[width=3.25in]{fig4.eps} 
  \caption[]{}
  \label{fig4}
\end{figure}

\clearpage

\begin{figure}  
  \centering \includegraphics*[width=5in]{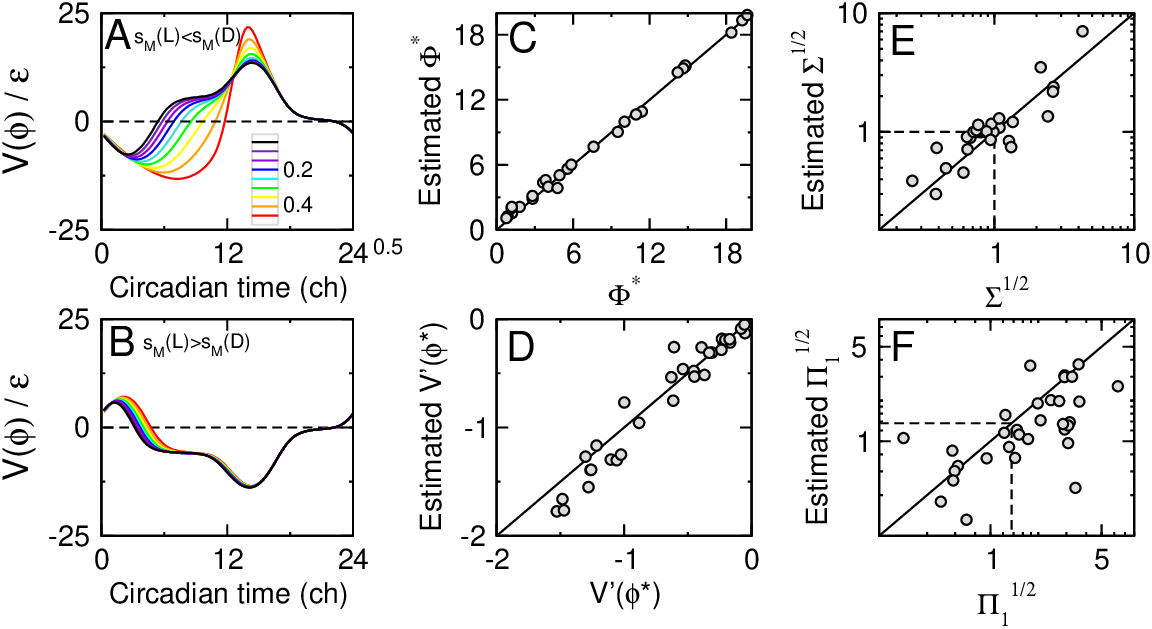} 
  \caption[]{} 
  \label{fig5}
\end{figure}

\clearpage

\begin{figure}  
  \centering \includegraphics*[width=4in]{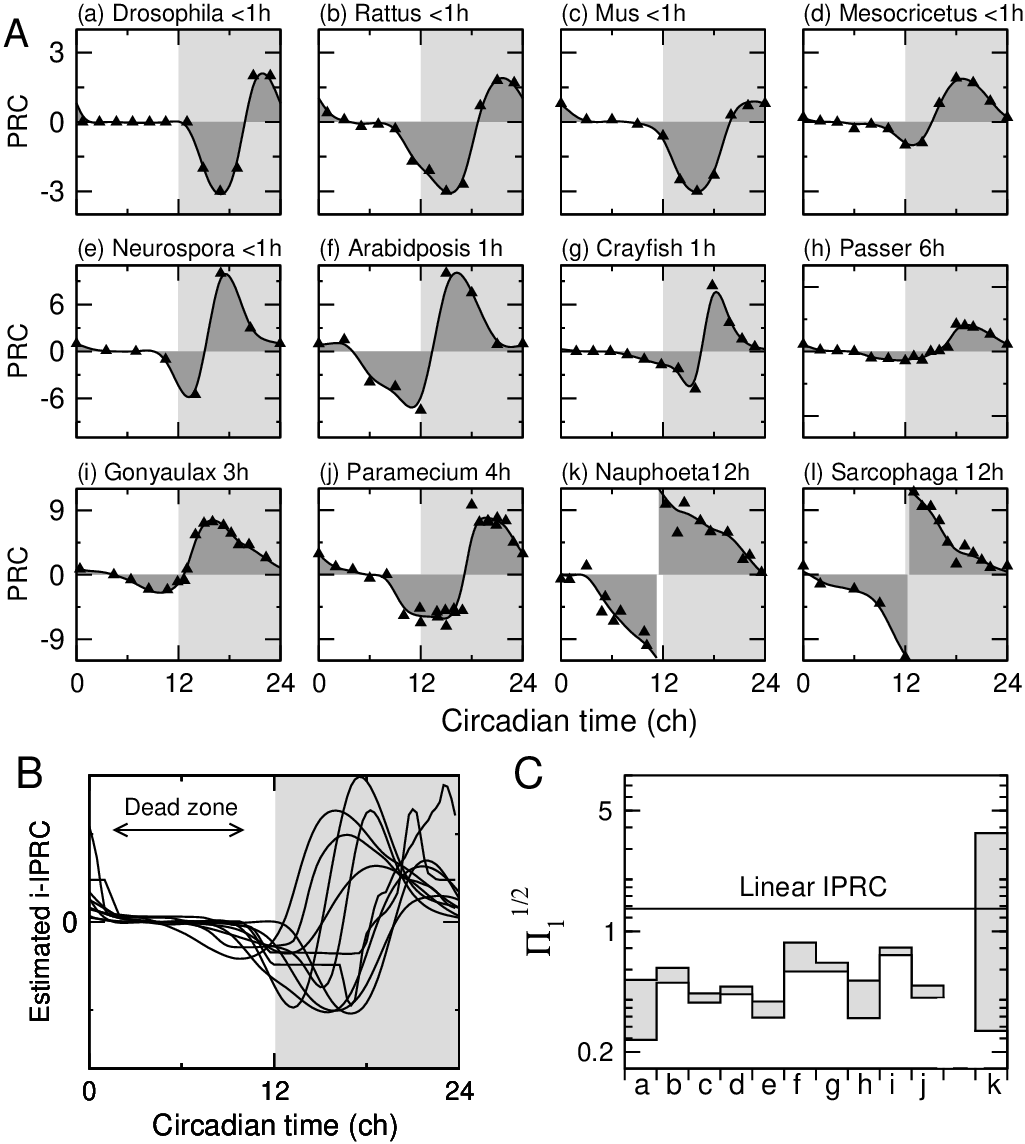} 
  \caption[]{} 
  \label{fig6}
\end{figure}

\end{document}